\documentclass[sigconf,screen,nonacm,pbalance]{acmart}
\author{
  Ferran Hermida-Rivera*,
  Jack Toubes*,
  Angela Cui,
  Chengyi Lux Zhang,
  Sagar Karandikar
}

\affiliation{%
  \vspace{1ex}
  \institution{
    University of California, Berkeley \\
    \vspace{1ex}
  }
  \city{}
  \country{}
}

\affiliation{\{ferran, jtoubes, angelacui, iansseijelly\}@berkeley.edu, sagark@eecs.berkeley.edu\country{}}

\thanks{\textsuperscript{*}Equal Contribution}

\copyrightyear{}
\acmYear{}
\acmDOI{}
\acmConference[]{}{}{}

\acmISBN{}

\usepackage{amsmath,amssymb,amsfonts}
\usepackage{algorithmic}
\usepackage{graphicx}
\usepackage{textcomp}
\usepackage{xcolor}
\usepackage{circuitikz}
\usepackage{pgfplots, pgfplotstable}
\usepackage{tabularx}
\usepackage{float}
\usepackage[font=small,labelfont=bf]{subcaption}
\usepackage{array}
\usepackage{layouts}
\usepackage{makecell}
\usepackage{placeins}
\usepackage{comment}
\usepackage{xurl}
\usepackage{hyperref}
\usepackage{tabularx,tabularkv,lipsum}
\usepackage{svg}
\usepackage{minted, listings}
\usepackage[font=small,labelfont=bf]{caption}
\usepackage[normalem]{ulem}
\useunder{\uline}{\ul}{}
\usepackage[normalem]{ulem}
\usepackage[autostyle, english=american]{csquotes}
\usepackage{enumitem}
\MakeOuterQuote{"}

\def\BibTeX{{\rm B\kern-.05em{\sc i\kern-.025em b}\kern-.08em
    T\kern-.1667em\lower.7ex\hbox{E}\kern-.125emX}}

\makeatletter
\newcommand{\thickhline}{%
    \noalign {\ifnum 0=`}\fi \hrule height 1pt
    \futurelet \reserved@a \@xhline
}

\makeatletter
\newcommand\blfootnote[1]{%
  \begingroup
  \renewcommand\thefootnote{}
  \protected@xdef\@thefnmark{}
  \@footnotetext{#1}
  \endgroup
}

\usepackage[colorinlistoftodos,prependcaption,color=yellow]{todonotes}

\definecolor{ClearSky}{RGB}{173, 237, 255}
\definecolor{Mist}{RGB}{232, 250, 255}
\definecolor{Cereal}{RGB}{255, 234, 150}
\definecolor{Wheat}{RGB}{255, 249, 223}
\definecolor{OceanDepth}{RGB}{26, 31, 161}
\definecolor{Abyss}{RGB}{15, 19, 96}
\definecolor{Spearmint}{RGB}{149, 251, 190}
\definecolor{Lichen}{RGB}{232, 253, 240}
\definecolor{Eraser}{RGB}{255, 195, 254}
\definecolor{RoseOfSharon}{RGB}{255, 233, 255}
\definecolor{HotPink}{RGB}{247, 0, 103}
\definecolor{HotBlue}{RGB}{13, 101, 255}
\definecolor{HotGreen}{RGB}{19, 180, 10}
\definecolor{HotOrange}{RGB}{255, 87, 51}
\definecolor{LightGray}{gray}{0.65}
\definecolor{MediumGray}{gray}{0.45}

\setcopyright{cc}
\setcctype{by}

\begin{document}

\newcommand{\pyinline}[1]{\lstinline[language=Python, basicstyle=\ttfamily, breaklines=true, breakatwhitespace=false]|#1|}

\title[BOOMerang: Autonomous Translation from RTL to Microarchitectural Simulators and Back]{BOOMerang: Autonomous Translation from RTL to Microarchitectural Simulators and Back}

\begin{abstract}
High-level microarchitectural simulators like gem5 are often used to test and demonstrate novel architectural techniques. While these simulators provide high research productivity, they generally lack the means to provide detailed feedback on the realizability of a hardware design. Furthermore, these simulators can frequently fall out-of-sync with the detailed performance characteristics of real hardware designs, which is usually the responsibility of dedicated engineers/teams in industry. While RTL implementation can address these challenges, it can be even more difficult to translate simulator code to high-quality RTL by hand without a large team. We demonstrate the ability to automatically address simulator-to-RTL alignment and translation challenges using automated agentic techniques.

In this work, we architect and evaluate agentic loops that automate both parts of the RTL vs. high-level simulator workflow that is critical to high-quality architecture research. Our loops (1) automatically generate accurate high-level models of RTL implementations of CPU designs (BOOM) in a battle-tested microarchitecture simulation framework (gem5) and (2) automatically translate new microarchitectural features implemented in the high-level simulator back to RTL faithfully.

Targeting the first loop at autonomously building a gem5 model for the Berkeley Out-of-Order Machine (BOOM) core, we generate a simulator with IPC within 3\% of the RTL, meeting targets pursued for simulators in industry, in only 2 days. We use our second loop to automatically translate 72 features from this aligned gem5 simulator into RTL in the BOOM core, while maintaining its low IPC error. Combining translated optimizations together yields a 25\% performance improvement for the BOOM core RTL implementation in SPECint 2017, demonstrating that having these loops for moving between high-level simulation and RTL (and vice-versa) enables actual improvement of processor performance. The agentic loops we developed and the simulators and RTL we generated will be available in open-source.

\end{abstract}
\maketitle

\section{Introduction}\label{sec:intro}
High-level microarchitectural simulators are commonly used by computer architects to explore novel ideas. Their ease of iteration and relatively concise codebases make them valuable tools in the hardware design process. While directly implementing ideas at the Register-Transfer Level (RTL) offers benefits in terms of fidelity and ASIC quality-of-result feedback, the relative difficulty of writing and verifying designs in hardware-description languages (HDLs) and the low throughput of RTL/HDL simulation make it difficult to rapidly explore novel ideas for early stage evaluation.

\begin{figure}[t]
    \centering
    \includegraphics[width=\columnwidth]{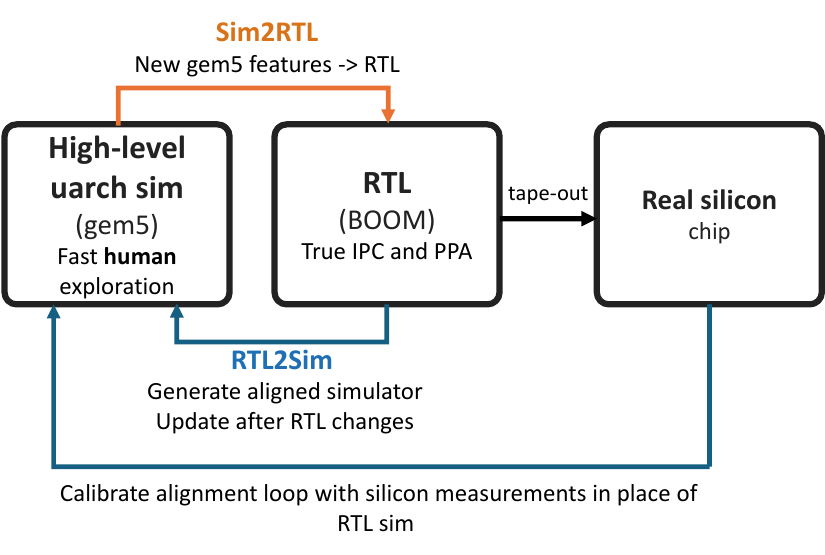}
    \caption{How Sim2RTL and RTL2Sim integrate into a computer architect's workflow.}
    \label{fig:arch-workflow}
\end{figure}

High-level simulator-based methodology relies on two assumptions: \textbf{(1)} that the high-level simulators are an accurate representation of the core they are intended to model and \textbf{(2)} the architect implementing features in these simulators has sufficient intuition about the underlying microprocessor and technology constraints to understand realizability outcomes, such as cycle-level timing impact and change in silicon area.

Unfortunately, there is no canonical or algorithmically automated way to align and translate between an RTL description of a processor and a corresponding microarchitectural simulator. For large teams, maintaining a simulator model of a processor has necessitated dedicating many engineer-hours to keeping the simulator up to date, validating its alignment with the true performance of the core, and translating new ideas back to RTL~\cite{OpenXiangShan22, alotoom2026}. For small teams, this is not feasible, and microarchitectural simulators fall out of alignment or are forgone completely. In practice, this means that the two assumptions above are often unreasonable ones, which may lead to incorrect research conclusions~\cite{Nowatzki15Harmful, Gibson00, Carlson14}.

In this paper, we design, build, and evaluate agentic AI-driven flows that automate both directions of this process, which we dub \emph{RTL2Sim} and \emph{Sim2RTL}, as shown in Figure~\ref{fig:arch-workflow}.
Our agentic translation and alignment system is built on CHIA~\cite{cui26}, which enables both re-use of our loops and the ability to re-target RTL2Sim and Sim2RTL to new simulators and new RTL implementations. To this end, our agentic system will be open-sourced in the near future.

Our RTL2Sim flow automatically generates and calibrates a microarchitectural simulator which accurately models an RTL design, and can keep the simulator up to date as the underlying design changes, even after the design is taped out. In this paper, we use RTL2Sim to automatically generate a highly performance-accurate gem5~\cite{gem5-new} simulator for the Berkeley Out-of-Order Machine (BOOM) core~\cite{Zhao20, Celio18}. In each iteration of the flow, an agent modifies the source code modeling the core in gem5 and configuration information to reduce differences between the behavior of the BOOM core's RTL and the existing gem5 simulator. After the agent finishes, its changes are thoroughly evaluated with a large suite of benchmarks, only some of which are visible to the agent. Each iteration builds off of the changes from a successful past iteration.

Our Sim2RTL flow enables automatically translating newly implemented features in the gem5 representation of a processor design \textit{into the BOOM RTL implementation}. These can either be brand new features implemented by an architect in gem5 or features assembled out of the vast library of microarchitectural techniques and blocks included in gem5. To perform the translation, in each iteration, an agent modifies the BOOM core's RTL to implement the novel feature so that the BOOM core's performance aligns with the gem5 simulator's. Once the agent is done, a Verilator~\cite{SnyderVerilator} simulator is built for the modified BOOM core, and a large suite of benchmarks is run to check whether the performance of the RTL matches gem5. These steps repeat until the two simulators report similar performance. Final evaluations of our system's translations include full runs of all 30 trillion instructions of the SPECint 2017 benchmark suite with reference inputs, demonstrating robustness.

We describe the evolution of the design of our agentic flows over time in response to experimental results. This provides instructive insight into the process of designing AI-driven hardware/software co-design flows and highlights strategies that are effective for improving agents' abilities to accomplish hardware-related tasks.

The final version of RTL2Sim builds a gem5 simulator which achieves a geomean absolute IPC error below 3\% of RTL simulation across a large, diverse set of training and hidden benchmarks. This is well within the targets pursued by developers of simulators for performance verification of commercial cores~\cite{alotoom2026, Moudgill99}, and matches the alignment of simulators like Panthera for the AmpereOne as represented in~\cite{alotoom2026}. Our final flow was able to produce these results, starting from the gem5 O3 simulator (a baseline with no relation to the BOOM core), in just two days.

To evaluate Sim2RTL, we use it to bring microarchitectural features from gem5 to the BOOM RTL. Simulators like gem5 make it easy to model new techniques and have been furnished with a great number of features and blocks that are not present on most open-source processors. 
Sim2RTL accurately translates 72 features into the RTL. Across the resulting implementations, RTL IPC is within 1.5\% of gem5 IPC on a large suite of training benchmarks. The translated features span across front-end, back-end and the memory system and vary widely in origin: some already exist in gem5 and need no changes, while others were implemented from scratch in gem5 by an LLM. Finally, we merge 8 of the translated features into a single core: on the full SPECint 2017 suite with reference inputs, it improves IPC by 28.9\% over baseline MegaBOOM, increasing critical path by only 3\%.

In summary, our contributions are as follows:
\begin{enumerate}[leftmargin=*]
    \item We develop two reusable agentic flows: RTL2Sim and Sim2RTL. The first builds a gem5 simulator with high fidelity to the underlying processor it models. The second accurately translates optimizations made in the gem5 simulator back into RTL to improve the performance of the real processor.
    \item We detail the lessons we learned from building our system about building effective and efficient agentic loops for hardware design tasks. 
    \item We develop a high fidelity gem5 simulator for the BOOM core, achieving performance parity within 3\% across a large and diverse suite of benchmarks.
    \item We accurately translate a wide set of optimizations of gem5 to the BOOM RTL. A merge of the top 8 of these translated features improves BOOM's wall-clock performance on SPECint 2017 by 25\%.
\end{enumerate}

Our agentic loops, generated simulator, and BOOM optimizations will be open-sourced in the near future.

\section{Background}\label{sec:BG}

Below, we highlight several key systems we harness to construct and demonstrate our AI-driven RTL2Sim and Sim2RTL methodologies: Chipyard, BOOM, FireSim, gem5, and CHIA.

\subsection{Chipyard, BOOM, and FireSim}

Chipyard~\cite{Amid20} is an open-source framework for designing, configuring, simulating, and evaluating systems-on-chip in an agile manner. It is widely used for implementation-based RISC-V SoC and microprocessor research because it integrates hardware generators, simulation tools, software toolchains, and implementation flows into a single environment. The Berkeley Out-of-Order Machine (BOOM)~\cite{Celio18,Zhao20}, included in Chipyard, is a silicon proven~\cite{Gonzalez21, CelioBroom19}, open-source, out-of-order superscalar RISC-V microprocessor generator implemented in the Chisel HDL~\cite{Backrach12}. We use BOOM in the context of a full SoC in Chipyard as the RTL design for which we build a microarchitectural simulator using RTL2Sim, and into which we translate features from the microarchitectural simulator using Sim2RTL. To validate the quality of our translations, we also harness FireSim~\cite{Karandikar18}, an FPGA-accelerated simulation platform, to run long-running workloads on our augmented RTL designs, such as SPEC with full reference inputs (over 30 trillion instructions).

\subsection{The gem5 Simulator}

gem5~\cite{gem5-new} is an open-source discrete-event computer architecture simulator. By combining micro-architectural and system-level simulation it enables users to simulate a wide variety of designs. Implementing microarchitectural features in gem5 cores is typically easier than in RTL, and simulations of gem5 cores are typically faster than RTL simulation. In this work, we build an RTL2Sim flow that produces a gem5 simulator that is well aligned to the RTL of the BOOM core by modifying not only parameters, but also implementations of differing microarchitectural features. We also take advantage of the large collection of features which have been implemented in gem5 over the years, but which have not percolated down into the RTL of open-source cores, to evaluate our Sim2RTL flow.

\subsection{Agentic Loops in CHIA}

Though coding agents like Anthropic's Claude Code~\cite{claudecode} and OpenAI's Codex~\cite{Codex} are growing increasingly capable of performing complex, long-horizon tasks as the models which back them improve, using these agents on their own can be very limiting. For example, using coding agents in isolation makes it difficult to ensure that certain verification/validation tasks have been completed, and that agents are isolated from modifying evaluation substrates to cheat.

Agentic loop engineering is the practice of constructing iterative processes that use AI agents (including coding agents such as Claude Code and Codex) to manipulate an artifact, evaluate that artifact outside of the AI agent's control, and feed generated metrics back to AI agents.

To build re-usable, scalable, and trustworthy agentic loops for Sim2RTL and RTL2Sim, we build our loops using the open-source CHIA project~\cite{cui26}. CHIA allows us to represent the Sim2RTL and RTL2Sim agentic loops we develop as executable graphs (e.g., Figures~\ref{fig:rtl2gem5-loop} and~\ref{fig:gem52rt-translation-loop}). Using CHIA, we can ensure that certain verification and validation steps are guaranteed to occur without the interference of the coding agent (e.g., trying to reward hack by modifying benchmarks). CHIA's flexible abstractions and large open-source library of nodes make it easy to adapt to experimental results by changing these loops. Additionally, CHIA enables us to do ``good science'' around flows, capturing them as re-usable artifacts as well as enabling fine-grained tracking of LLM model/coding agent costs, compute costs, and time spent in various parts of loops. Even with this scaffolding, within a CHIA coding agent node, the coding agents are allowed to employ their full frontier capabilities, which scale with model quality over time.
Our Sim2RTL and RTL2Sim loops will be open-sourced as CHIA loops at publication time, providing fundamental tools for the research community to use.

\section{A Reusable Agentic Loop for RTL2Sim}\label{sec:RTL2Gem5}

\begin{figure}[t]
    \centering
    \includegraphics[width=1\columnwidth]{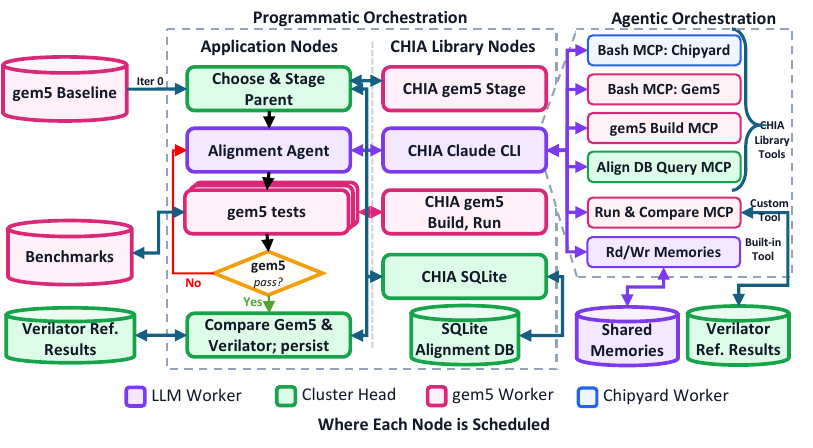}
    \caption{RTL2Sim: an agentic loop for automatically generating a representative gem5 microarchitectural simulator from RISC-V BOOM RTL.}
    \label{fig:rtl2gem5-loop}
\end{figure}

As mentioned above, high level simulator-based computer architecture research relies on the simulator being performance-accurate relative to the underlying RTL that it is simulating. However, building these accurate simulators, keeping them up to date as RTL changes, and calibrating them with real-silicon after taping-out a design is a huge engineering task. Figure~\ref{fig:rtl2gem5-loop} details RTL2Sim, the agentic loop we designed to generate, calibrate, and keep up-to-date a microarchitectural simulator which accurately models the performance of an RTL implementation of a processor. We target RTL2Sim at generating a gem5 simulator for the BOOM core. In this section we detail the design of RTL2Sim and the lessons we learned while building it.

At the highest level, each iteration of the loop consists of four parts (\textit{application nodes} in Figure~\ref{fig:rtl2gem5-loop}). First, a parent gem5 simulator (the result of a past iteration of the loop) is chosen and copied (staged) into a gem5 codebase. Next, using prompting and tools, an agent is provided access to and information about both this parent simulator which it will be asked to incrementally improve, and the RTL of the BOOM core it is basing the gem5 simulator on. It is also provided the results from simulating a set of \textit{training} benchmarks in both the parent iteration's gem5 simulator and software RTL simulation (Verilator~\cite{SnyderVerilator}). The agent is told that its target is to improve the gem5 simulator such that its IPC more closely matches the core's IPC as measured in RTL simulation, and is then asked to modify the gem5 simulator's underlying source code and configuration accordingly.

Third, after the agent finishes, its improved gem5 simulator is built and used to simulate both the aforementioned training benchmarks, as well as a collection of hidden \textit{holdout} benchmarks (described in more detail in Section~\ref{sec:rtl2gem5:bmarks}). Finally, the difference in IPC between the gem5 simulations and the RTL-level simulations is calculated, and the results are saved. 

\begin{figure}[ht]
    \centering
    \includegraphics[width=\columnwidth]{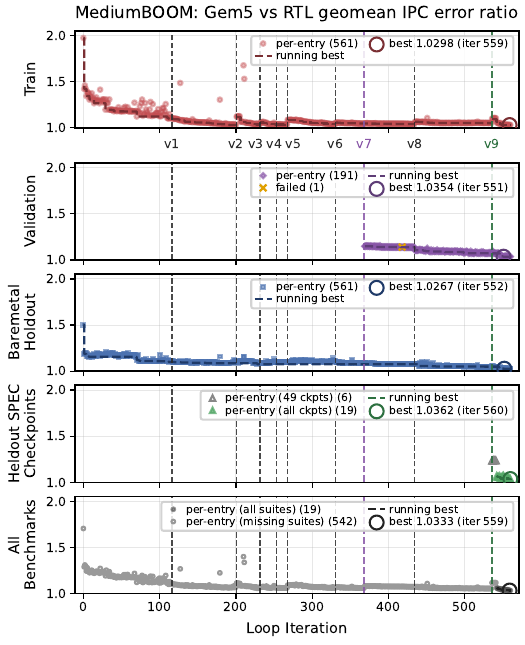}
    \caption{Absolute IPC error ratio between gem5 and RTL simulation in Verilator for each iteration of the gem5 simulator generated by the agents as RTL2Sim was being developed. Lower is better, 1 is perfect alignment between the simulations. Vertical dotted lines indicate significant changes to the loop.}
    \label{fig:rtl2gem5-conv-exp}
\end{figure}

Though this four-phase structure of RTL2Sim remained consistent, as our experiments progressed, the loop evolved to solve a number of problems and inefficiencies which we encountered. In Figure~\ref{fig:rtl2gem5-conv-exp}, we show the results of our experiments aligning the Medium (2-wide) configuration of the BOOM core's RTL with a gem5 simulator, with clear demarcations indicating when changes were made to the loop, as described in the text below. The metric shown in Figure~\ref{fig:rtl2gem5-conv-exp} is absolute IPC error ratio, and is defined below in Equation~\ref{eqn:absolute-IPC-error}. In Section~\ref{sec:eval-RTL2Gem5}, we evaluate the final version of RTL2Sim when run with the improvements discussed below.

\begin{equation}
    geomean_{bmarks}\left(max\left(\dfrac{IPC(gem5)}{IPC(RTL)}, \dfrac{IPC(RTL)}{IPC(gem5)}\right)\right)  \label{eqn:absolute-IPC-error}
\end{equation}

\subsection{Agent Knowledge and Capabilities}\label{sec:rtl2gem5:capabilities}

In addition to the gem5 source and the processor's RTL, from the beginning of our experiments we also provided the agent access to the source and disassembly of the experiment's training benchmarks (distinct from hidden benchmarks, as mentioned above), as well as instruction traces and performance counters from the parent iteration's gem5 simulations and the RTL-level simulations of the training benchmarks. To allow the agents to learn from each other, especially agents with different lineages of parent simulators, the agents were also given a tool to allow them to query the database of past agent attempts. Though a selection of this information was included in the initial prompt for the agent, an emphasis was placed on making information available primarily through agentic tool calling, allowing the agent to bring information into context when it considered it useful. When examining logs, we saw pipeline traces leading to identification of misaligned structures and database access leading to merges of multiple past successful improvements.

However, this experimental setup led to a lot of regressing iterations, where IPC alignment got worse. To address this, at the line marked "v1" in Figure~\ref{fig:rtl2gem5-conv-exp}, we gave the agents the ability to test their changes, by providing tools which an agent could call to compile the gem5 simulator it had generated, and run it on any training benchmarks. After this addition, we saw the agents begin to run their own experiments, collect evidence empirically, and use it to change their simulator (often multiple times) before exiting and sending it back for more thorough evaluation.

\subsection{Benchmarking Strategy}\label{sec:rtl2gem5:bmarks}

In line with other work developing accurate high-level simulators for processor cores~\cite{zhen2026, alotoom2026}, our original suite of training benchmarks (top panel in Figure~\ref{fig:rtl2gem5-conv-exp}) consisted of 36 microbenchmarks designed to stress individual components of an out of order core, selected from Vertical Research Group's MICROBENCH suite \cite{Nowatzki15}. In addition, we started with a hidden, holdout suite comprised of 42 baremetal benchmarks from Embench~\cite{Patterson25}, CoreMark-PRO~\cite{eembc2015}, and MiBench~\cite{Guthaus01}. Benchmarks duplicated in multiple of these suites were only included once. We used the hidden benchmarks to help guide our experimentation by measuring how well the alignment between the gem5 simulator and the actual RTL generalized to benchmarks which could not be seen by the agents. The results on this suite were never made visible to the agents, even indirectly (the results on the hidden suite were \textit{not} used for parent selection for example). As we ran the experiment, it became clear quickly that performance accuracy of the gem5 simulator for the training benchmarks was improving continuously, but for the holdout benchmarks it quickly plateaued. The IPC alignment for this original holdout suite is shown in the third panel in Figure~\ref{fig:rtl2gem5-conv-exp}. 

To address this, we progressively made additions to our benchmark suite with the goal of forcing the agents to make more generalizable changes to the gem5 version of the core. First, we tried using agents to generate benchmarks which were designed to highlight misalignments. Over 4 rounds of benchmark generation (inserted at the lines marked "v2" through "v5" in Figure~\ref{fig:rtl2gem5-conv-exp}) plus some iterations of RTL2Sim, 50 benchmarks were generated and added to the training suite (for a total of 86). Though the agents did produce adversarial benchmarks (the benchmarks they generated decreased alignment, as seen by small jumps in the training suite's mis-alignment at lines "v2" through "v5"), the alignment agents did not produce more generalizable gem5 simulators, as the holdout alignment did not improve.

Our next strategy was introducing a set of \textit{validation} benchmarks (inserted at the line marked "v7" in Figure~\ref{fig:rtl2gem5-conv-exp}). Alignment convergence is shown for these validation benchmarks in the second panel of Figure~\ref{fig:rtl2gem5-conv-exp}. Inspired by ML model training, these validation benchmarks were invisible to the alignment agents \textit{except} for being used for parent iteration selection (together with the training suite). We selected 44 benchmarks for our validation suite from TACLeBench~\cite{Falk16}, BEEBS~\cite{Pallister13}, and the LLVM test-suite~\cite{llvmtestsuite20}, ensuring that all 44 were not in the existing holdout set. 

Upon adding the validation set to RTL2Sim, we began to see some movement in our holdout suite's alignment, but it was still mostly flat. For this reason, we decided to try moving half (22) of our validation benchmarks into the training suite (for a total of 108 training benchmarks), since these benchmarks were more structurally similar to our holdout benchmarks than our existing training benchmarks (longer, more varied workloads as opposed to structure-targeted microbenchmarks). We made this change at the line marked "v8" in Figure~\ref{fig:rtl2gem5-conv-exp}. Upon making this change, we finally began to see a significant amount of improvement in the performance-accuracy of our gem5 core relative to the RTL for the holdout suite. From this, we conclude that it was critical for the training suite to contain benchmarks which were more structurally representative of the benchmarks in the holdout set.

Finally, to cover the effects of running an operating system, at the line marked "v9" in Figure~\ref{fig:rtl2gem5-conv-exp}, we converted some of our training benchmarks to be run in Linux, and added a second hidden (heldout) set of benchmarks composed of 130 SimPoints~\cite{Sherwood02} from the SPECint 2017 benchmark suite~\cite{Bucek17}. We also take one of the SPECint benchmarks' set of SimPoints out of the holdout suite and put it in the training suite. This added 12 more benchmarks to the training suite for a total of 120. The IPC alignment for the heldout SimPoints is shown in the fourth panel in Figure~\ref{fig:rtl2gem5-conv-exp}.

\subsection{Evaluation Metrics and Parent Selection}

In RTL2Sim, each iteration begins with choosing and restaging a parent iteration from which to further iterate. Our first strategy for choosing this parent was to take just the most closely aligned existing iteration in the database. This worked well towards the beginning of our experiments, as the low-hanging fruit alignment opportunities (e.g. structure sizes) generally improved alignment by significant amounts and across all benchmarks.

When we added LLM experimentation, we switched to choosing randomly between the three iterations most closely aligned to the BOOM core across the suite on average. As mentioned in Section~\ref{sec:rtl2gem5:bmarks}, we later introduced a validation suite which was used in tandem with the training suite to determine which iteration was most closely aligned to the BOOM core.

At the line marked "v6" in Figure~\ref{fig:rtl2gem5-conv-exp}, we also introduced a set of top-down microarchitectural analysis~\cite{Yasin14} (TMA) counters into gem5, designed to be similar to the ones in BOOM~\cite{Weingarten25}, though subject to alignment (modification) by the agents just like the rest of the core. After introducing these counters, we switched to choosing parents from a Pareto-frontier of entries, which were Pareto-optimal with one axis being TMA attribution misalignment and the other being IPC misalignment. 

After making this change, we did not see a significant improvement in IPC alignment, though our TMA attribution alignment did improve significantly, implying the core was becoming more representative. When we introduced the TMA counters at "v6", the share of issue slots that gem5 attributed to a different top-level bottleneck than the RTL was 8.37\% on average across the training benchmarks. This share was reduced to 3.27\% by the end of our experiments. The top-level bottleneck buckets are retiring instructions, frontend bottlenecked, backend bottlenecked, and bottlenecked by bad speculation. We do not claim these results to be truly indicative of alignment, since the agents have the freedom to change the computation of the TMA counters in the gem5 simulator, making it an unreliable metric.

\subsection{Model Selection}

When we began our experiments, the state-of-the-art frontier model was Claude Opus 4.6, and this is what we used for the majority of our experiments. We experimented with later versions of Opus 4 and with Claude Fable 5, but found that they did not yield a significant improvement in alignment in each iteration, despite increases in cost. At the line marked "v9" in Figure~\ref{fig:rtl2gem5-conv-exp}, we switched Opus 4.6 for Opus 5.5 which did improve the loop trajectory. The lessons described in this section for improving the performance of the models on this task are thus most relevant for models with the same level of capability as Claude Opus 4.6.

\subsection{Loop Design Principles}

In summary, we derive the following lessons from our experiments designing RTL2Sim.

\begin{itemize}[leftmargin=*]
    \item A shared database of knowledge unlocks compounding benefits from many parallel iterations of an agentic loop.
    \item Providing the agents with tools to evaluate their changes empirically increases the likelihood that each iteration improves a design.
    \item A substantial evaluation which the agent cannot see (e.g. our withheld benchmarks) provides important feedback on the generality of the results.
    \item On the other hand, the withheld evaluation must be well-enough represented in the training suite for the agent to achieve good results.
    \item Evaluating models as they are released is critical to ensure that an agentic loop is using the best available agent for the job. 
\end{itemize}

\section{A Reusable Agentic Loop for Sim2RTL}\label{sec:Gem52RTL}

Given the difficulty of implementing new microarchitectural ideas in an RTL implementation of a microprocessor, many new ideas are evaluated exclusively in high-level microarchitectural simulators, with the vast majority of more detailed evaluation happening behind closed doors in industry.
As a result, key questions remain open for these ideas: whether they are physically realizable, their real area and power impact, and whether their IPC gains translate to the constraints of a real RTL implementation of a core in a complete system-on-chip~\cite{Nowatzki15Harmful, Gutierrez14}.

\begin{figure}[t]
    \centering
    \includegraphics[width=\columnwidth]{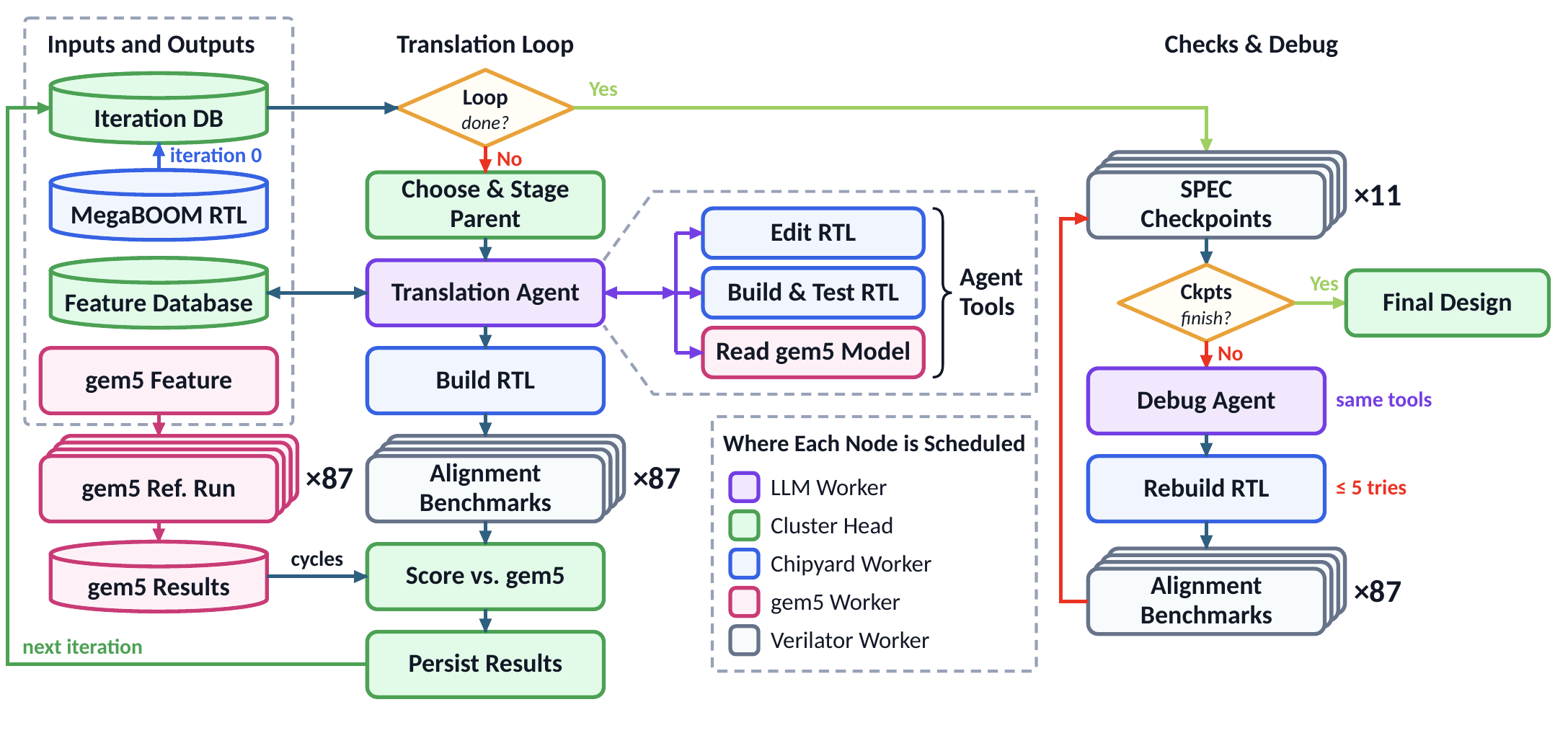}
    \caption{Diagram of Sim2RTL, our agentic loop used to translate a given feature down to RTL, MegaBOOM's RTL in this case.}
    \label{fig:gem52rt-translation-loop}
\end{figure}

To address this, we developed Sim2RTL, an agentic loop capable of translating features from a C++ gem5 simulator implementation to RTL implementations in the BOOM core.
Figure~\ref{fig:gem52rt-translation-loop} depicts the final version of Sim2RTL. We start with a gem5 simulator that has been previously aligned to our target RTL (the product of the RTL2Sim loop in Section~\ref{sec:RTL2Gem5}). Some entity (e.g., a human architect or agent) then augments the gem5 simulator with a new microarchitectural feature, either built from existing gem5 primitives, or coded from scratch (e.g., for a completely new idea). 
Introducing the new feature naturally induces misalignment between the gem5 simulator and the RTL. Therefore, the high-level task of Sim2RTL is to realign them by implementing, in the RTL, a faithful representation of the newly added feature, and fixing existing microarchitectural mismatches that may be exposed by the new feature, as well as debugging the final implementation to ensure it is fully functional.

Like the RTL2Sim loop in Section~\ref{sec:RTL2Gem5}, this task is performed iteratively; each iteration builds off of a prior iteration's RTL. The translation agent is given a natural language description of the feature it will translate, and a tool to read the gem5 implementation of the feature. The LLM is instructed to treat the gem5 source code as the ground-truth specification for the feature. It also receives, for each of the benchmarks in the train suite, its source code, a short description, and a full set of performance counters collected in both gem5 and Verilator simulations of the benchmark. The LLM also has access to a small database where prior iterations have logged useful findings and information, such as what parts have not yet been translated and analyses of unusual results. With all of this information, and a tool to test its changes in Verilator simulations, the agent is asked to perform the translation. It is also encouraged to add new counters to profile the added feature and to identify the remaining gaps upon finishing each iteration. Then, it logs everything into the feature database.

The next step in each iteration of Sim2RTL consists of running a set of training benchmarks on the resulting RTL in Verilator. Upon finishing these simulations, the loop computes the mean absolute cycle difference between gem5 and RTL, and logs both the designs and the score in the iteration database. The loop then begins a new iteration or stops if the score has reached the desired threshold. Upon reaching this threshold, the final RTL is tested on a set of much heavier benchmarks, SPEC17 checkpoints, and debugged autonomously in case any problems arise.

The version depicted in Figure~\ref{fig:gem52rt-translation-loop} and described above is the final version and the one we use throughout the evaluation in Section~\ref{sec:eval-Gem52RTL}. 
We describe many of our design decisions in more detail below.

\subsection{Benchmarks and Metrics}\label{sec:gem52rt-sub-loop}
Sim2RTL measures alignment on 87 bare-metal benchmarks drawn from the training suite presented in Section~\ref{sec:rtl2gem5:bmarks}. Specifically, we take the 87 shortest bare-metal tests. We restricted the loop to these benchmarks to cut down the iteration time, given that Verilator simulations now execute in-loop in each iteration (as opposed to the significantly faster gem5 simulations run in-loop in Section~\ref{sec:RTL2Gem5}).

To compute performance-alignment for each of the 87 benchmarks, Sim2RTL measures the cycle count of both gem5 and the RTL, and computes the absolute error $e_b = |C^{\mathrm{RTL}}_b - C^{\mathrm{gem5}}_b| / C^{\mathrm{gem5}}_b$. The score is the mean of $e_b$ over all 87 benchmarks, where a benchmark with $e_b < 1\%$ counts as $e_b = 0$. The reason for this rule is that a benchmark that is within 1\% of error should be considered aligned and small changes in it should not prevent meaningful change that fixes a clear gap from being accepted. The loop terminates when a design hits $score \leq 1.5\%$.

\subsection{Agent Knowledge and Capabilities}\label{sec:gem52rtl-sub-agents}

The loop normally uses the score presented in Section~\ref{sec:gem52rt-sub-loop} to choose the design that the next iteration starts from. In earlier versions of Sim2RTL, however, we found that a new feature can expose previously hidden misalignments between the gem5 baseline and the RTL, so the most faithful design does not always score best. In these cases, we want the agent to fix the exposed misalignment rather than abandon the design or alter the feature to hide it. We therefore let the agent mark a design with a \texttt{KEEP} flag when it believes the score penalizes the design only because of such a mismatch; the next iteration then starts from the flagged design regardless of its score. We have seen that this rule, combined with treating the gem5 implementations as the true specification helps improve the quality of translation.

Distinguishing cycle mismatches caused by an inaccurate feature implementation from those arising from other misalignments has often proven challenging for the agent. To address this issue, we introduced a new rule requiring the LLM to implement counters that profile the feature's behavior. These counters provide additional insights that help the agent make this distinction, ultimately improving the overall quality of the generated RTL changes.

Implementing a faithful translation and fixing the misalignments exposed takes many implement and test rounds within one iteration, as well as many overall iterations. We see agentic turns taking several hours and performing hundreds of tool calls, leading to huge contexts that are eventually compacted. To mitigate this, we implemented a per-feature database in which, after each iteration, the agent stores relevant information for future iterations. This database also offers robustness against failures and rate-limit errors because each iteration can be allocated a new fresh agentic session without losing valuable information.

\subsection{Loop Design Principles}
While developing Sim2RTL we derived the following lessons.
\begin{itemize}[leftmargin=*]
    \item A way to temporarily keep a design that is not the best-scoring one (the \texttt{KEEP} flag) reduces compensating changes that do not match the model.
    \item Sharing a condensed set of learnings and the remaining work across iterations helps the agents stay on track.
    \item Obtain a stronger signal on the alignment of the feature by telling the agent to explicitly profile it (e.g. adding counters) helps faithfulness.
    \item Not attending to the benchmarks whose alignment variations are well within the alignment goal (e.g. 1\%) help the loop to rewards meaningful changes.
\end{itemize}
All the runs of our final loop have been made with Opus 5.5.

\section{Evaluation: RTL2Sim}\label{sec:eval-RTL2Gem5}
As mentioned in Section~\ref{sec:RTL2Gem5}, we targeted RTL2Sim at generating a high-level simulator in gem5 for the BOOM core. We collected results as we developed the loop, starting from the generic gem5 RISC-V O3 core model as our baseline. Figure~\ref{fig:rtl2gem5-conv-exp} shows the absolute IPC error ratio (Equation~\ref{eqn:absolute-IPC-error}) between the gem5 simulator and the RTL for RTL2Sim as it was developed. In this section, we describe results from evaluating the \textit{final}, fully-developed version of RTL2Sim. In the results described below, the IPC of the RTL was measured using Verilator with an idealized low-latency DRAM model, and except where otherwise stated below, the loop was targeted at the Medium (2-wide) configuration of the BOOM core.

\begin{figure}[t]
    \includegraphics[width=\columnwidth]{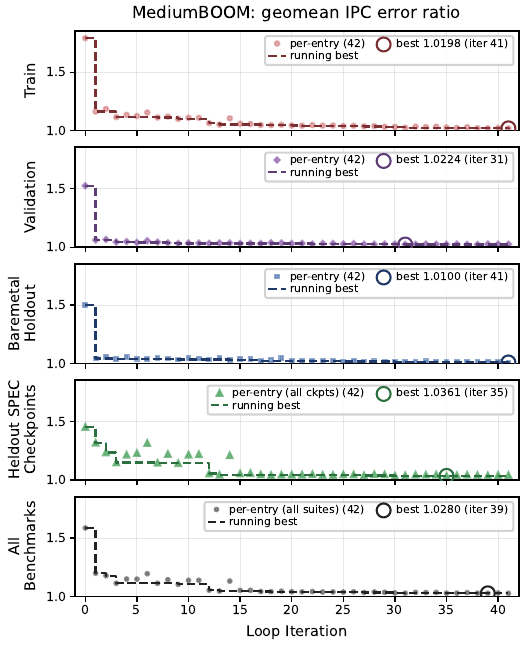}
    \caption{Absolute IPC error ratio between gem5 simulator and RTL simulation in Verilator for each iteration of the gem5 simulator generated by the agents (with Opus 5.5) in the final version of RTL2Sim. Lower is better, 1 is perfect alignment between the simulations.}
    \label{fig:rtl2gem5-conv-eval}
\end{figure}

To evaluate RTL2Sim, we re-ran the final, most successful version of the loop, starting again from the baseline. Figure~\ref{fig:rtl2gem5-conv-eval} shows the absolute IPC error ratio between the gem5 simulator and the RTL as the gem5 simulator evolved, broken down by benchmark suite (these suites are described in Section~\ref{sec:rtl2gem5:bmarks}). This version of the flow converges to a very low misalignment: the loop achieves a geomean IPC ratio compared to RTL simulation of the BOOM core of only 1.0280 (within 3\%). On the 130 held-out SimPoints of SPECint 2017, the geomean IPC ratio is 1.0361. 96\% of benchmarks are within 10\%, including 98.3\% of training benchmarks. These results are situated within targets pursued by developers of simulators for performance verification of commercial cores~\cite{alotoom2026, Moudgill99}, including being far beyond Ampere Computing's "goal of having 90\% of tests within ±10\%" IPC~\cite{alotoom2026}, where all benchmarks are visible to the engineers implementing the simulator (akin to our training suite).

\begin{figure}[t]
    \centering
    \includegraphics[width=\columnwidth]{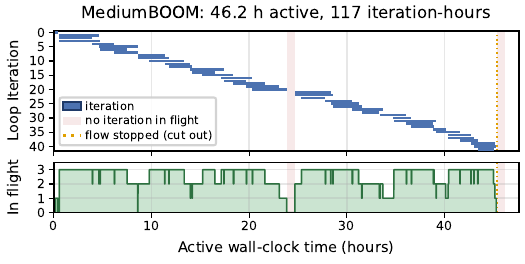}
    \caption{\textbf{Top:} Wallclock run time of each iteration of the final version of RTL2Sim. 41 iterations of the loop were run over a two day span. \textbf{Bottom:} Parallel iterations of the loop over the duration of its run.}
    \label{fig:rtl2gem5-runtime-eval}
\end{figure}

Figure~\ref{fig:rtl2gem5-runtime-eval} shows the wallclock runtime of each iteration of RTL2Sim. As shown, the loop achieved this very-low misalignment in less than 2 days. The loop ran with a maximum parallelism of 3 iterations at a time, and finished 41 iterations beyond the baseline, with an average iteration time of 2.85 hours. 

We believe this fast convergence can be attributed to three things. \textbf{(1)} The final version of RTL2Sim is more effective than the loop was as it was being developed. \textbf{(2)} Any bad decisions which were made by the iterations run in a primitive version of RTL2Sim had to be corrected by later iterations, whereas the re-run did not have to dig out of these types of holes. \textbf{(3)} Opus 5.5 can perform more work in a single iteration than Opus 4.6 was able to.

\begin{figure}[t]
    \centering
    \includegraphics[width=\columnwidth]{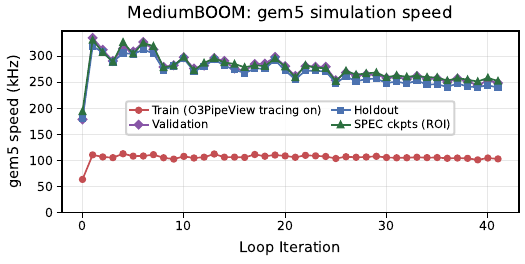}
    \caption{gem5 simulator speed in thousands of cycles per second for each iteration of the final version of RTL2Sim. Higher is better.}
    \label{fig:rtl2gem5-speed-eval}
\end{figure}

Table~\ref{tab:mediumboom-llm-cost} shows the LLM token costs and API-equivalent dollar costs spent to run the loop. Figure~\ref{fig:rtl2gem5-speed-eval} shows the speed of the gem5 simulator on each of the benchmark suites over the course of the evaluation. Relative to the baseline (iter 0) gem5 RISC-V O3 core (which has no relation to Medium BOOM), the Medium BOOM accurate simulator is significantly faster. However, as RTL2Sim makes the simulator more accurate, the simulation speed does progressively decrease, dropping from a peak around 335kHz to around 250kHz for the final iteration (around a 25\% drop).

\begin{table}[t]
\centering
\caption{Total and average LLM token consumption and cost for the fully-developed loop for generating a high-fidelity gem5 core for modeling Medium BOOM. Results shown for both Claude Opus 5.5 (see Figure~\ref{fig:rtl2gem5-conv-eval}) and Qwen 3.8 27B (see Figure~\ref{fig:rtl2gem5-conv-qwen}).}
\label{tab:mediumboom-llm-cost}
\label{tab:mediumboom-qwen-llm-cost}
\resizebox{\columnwidth}{!}{
    \begin{tabular}{ll*{6}{>{\raggedleft\arraybackslash}p{1.2cm}}}
    \hline
    Model & & LLM Iters & Input Tokens & Cache Creation & Cache Read & Output & Cost (\$)\\
    \thickhline
    Opus 5.5 & Total     & 41 & 21.0K & 30.7M & 5.13B & 14.4M & 1.56K \\
             & Per iter. &  1 &   513 &  749K &  125M &  351K & 38.1  \\
    \hline
    Qwen 3.8 27B & Total     & 38 & 65.1M & 0 & 677M  & 6.16M & 44.6 \\
                 & Per iter. &  1 & 1.71M & 0 & 17.8M &  162K & 1.17 \\
    \hline
    \end{tabular}
}
\end{table}

\begin{figure}[t]
    \centering
    \includegraphics[width=\columnwidth]{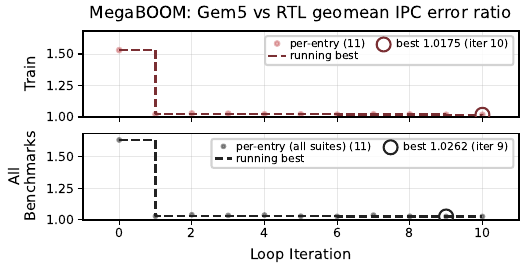}
    \caption{Absolute IPC error ratio between gem5 simulator for MegaBOOM, and RTL simulation in Verilator for each iteration of the gem5 simulator generated by the agents (with Opus 5.5), starting from the best generated Medium BOOM simulator at iteration 0. Lower is better, 1 is perfect alignment between the simulations.}
    \label{fig:rtl2gem5-conv-mega}
\end{figure}

To test whether the task would be significantly more difficult for a larger configuration of the BOOM core, and to seed the work in Section~\ref{sec:eval-Gem52RTL}, we took the well aligned gem5 simulator for the Medium BOOM configuration, and had RTL2Sim retarget the simulator at the Mega (4-wide) configuration of the BOOM core. The two configurations differ only in the sizes of structures, as well as a small number of small changes to the underlying implementations of certain structures. The results of this retargeting are shown in Figure~\ref{fig:rtl2gem5-conv-mega} for the training benchmark suite, and all benchmarks (training, validation, baremetal holdout, and SPEC2017 SimPoint holdout). RTL2Sim clearly has no trouble with the larger configuration, converging in a single iteration. Furthermore, this experiment demonstrates the ability of RTL2Sim to quickly update the simulator after incremental RTL changes.

\begin{figure}[t]
    \centering
    \includegraphics[width=\columnwidth]{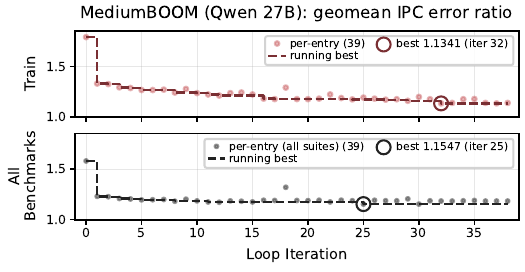}
    \caption{Absolute IPC error ratio between gem5 simulator and RTL simulation in Verilator for each iteration of the gem5 simulator generated by the agents (with Qwen 3.8 27B) in RTL2Sim. Lower is better, 1 is perfect alignment between the simulations.}
    \label{fig:rtl2gem5-conv-qwen}
\end{figure}

To further stress test RTL2Sim and its flexibility, we also run the final version of the loop for 2 days using the very small open-weight, Qwen 3.8 27B model. This model is small enough that it can be run on a laptop. These results are shown in Figure~\ref{fig:rtl2gem5-conv-qwen}. Though the results are not as strong as those achieved with Opus 5.5, they are decent, turning a completely unrelated baseline simulator into one which is somewhat aligned with a 1.1547 geomean IPC ratio (eqn~\ref{eqn:absolute-IPC-error}, about 15 percent off) at a total cost of only \$44.6, as seen in Table~\ref{tab:mediumboom-qwen-llm-cost}. This demonstrates that well structured agentic loops can accomplish decent results with small models.

\section{Evaluation: Sim2RTL}\label{sec:eval-Gem52RTL}

In this section we evaluate Sim2RTL by deploying it in a broader agentic loop we construct to translate a large number of features from gem5 into the RTL implementation of the BOOM core as part of a complete Chipyard SoC.

\subsection{Sim2RTL Outer Evaluation Loop}\label{sec:gem52rtl-eval:eval-loop}

\begin{figure*}[t]
    \centering
    \includegraphics[width=0.9\textwidth]{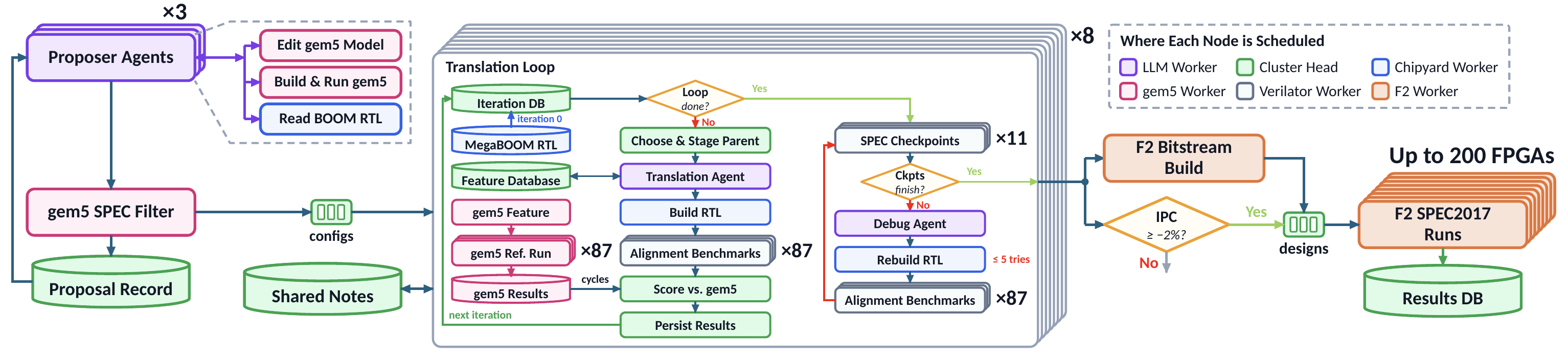}
    \caption{CHIA loop used to propose and translate many features in parallel.}
    \label{fig:gem52rt-distillery-loop}
\end{figure*}

We designed an ``outer'' agentic loop depicted in Figure~\ref{fig:gem52rt-distillery-loop} in order to evaluate the performance of Sim2RTL. This is divided into three main stages.

\textit{Stage 1: }First, a set of agents look for missing features in BOOM that already exist in gem5 or could easily be implemented. We ensure the agents propose unique suggestions with a shared database of proposed ideas, and by targeting each agent at a different part of the core (front-end, back-end, and memory system). Each proposed gem5 feature is evaluated on a representative set of checkpoints of the SPECint 2017 benchmark suite~\cite{Bucek17} to measure their IPC in gem5. We discard features that degrade performance by more than 0.5\%. These  checkpoints are referred to as \textit{SPEC checkpoints} for the rest of this section.

\textit{Stage 2: } The second stage is the main part of the loop. It is composed of 8 parallel instances of Sim2RTL, each of them translating one proposed feature at a time, and running through the whole flow described in Section~\ref{sec:Gem52RTL}. The translation agents can store notes and session transcripts on disk (\textit{shared notes} in Figure~\ref{fig:gem52rt-distillery-loop}) so later sessions can learn from earlier sessions. All alignment results from Sim2RTL are reported as geomean absolute IPC error, see Eqn~\ref{eqn:absolute-IPC-error}, though mean absolute cycle error was used by the loop.

\textit{Stage 3: }Lastly, all features are evaluated on the SPEC checkpoints for final alignment and performance improvement over the baseline BOOM core. Critically, for the first 36 designs that passed the SPEC checkpoints, we also build an FPGA bitstream and run the full SPECint 2017 intspeed suite with reference inputs in FireSim~\cite{Karandikar18} (30+ trillion instructions). We will refer to this particular suite as SPEC17 ref. throughout this section. Due to time and FPGA capacity constraints, we could not run all designs in FireSim.

\subsection{Sim2RTL Translation Alignment and Correctness}\label{sec:gem52rtl-eval:Results}

 \begin{figure}[t]
    \centering
    \includegraphics[width=\columnwidth]{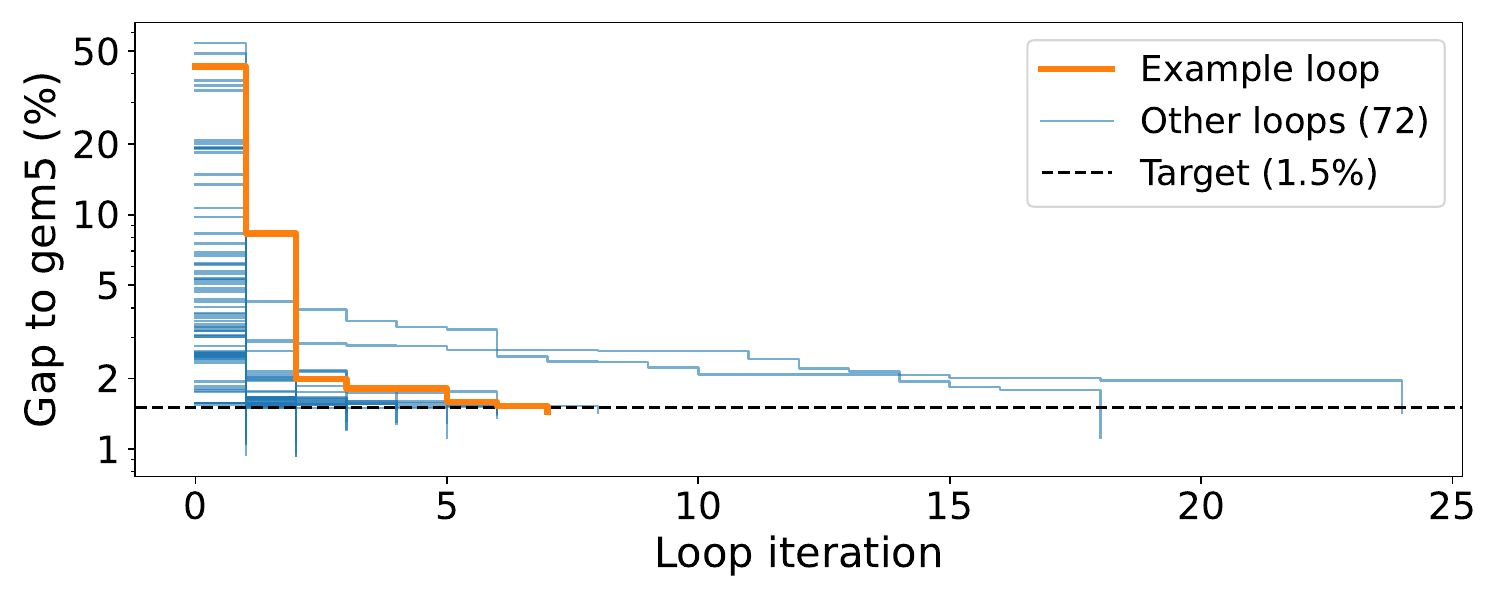}
    \caption{Alignment of the 73 translated features per iteration: each blue line is the best gap to gem5 so far
    (mean cycle difference in the training suite) of one loop. Dashed line is the 1.5\% target.}
    \label{fig:loop-alignment}
  \end{figure}
  
Over a span of 10 days, the evaluation loop proposed 74 features, of which 73 improved the performance of the gem5 representation of BOOM and made it to the translation step. Sim2RTL successfully translated 72 of them, and one had a bug which could not be fixed by the debugging agent.
Figure~\ref{fig:loop-alignment} shows the score of the translated features across iterations and highlights one of them for visualization purposes. 
On average, each feature took 2.7 iterations to translate and the average feature has an absolute IPC error ratio of 1.015 on the train set, with the worst translation achieving 1.017, better than the original alignment of 1.018. On the checkpoints run in stage 3, the average ratio across all features is a bit higher, but still quite low at 1.048, with a maximum of 1.060. 

 Table~\ref{tab:feature-breakdown} shows how the features are distributed across the core's RTL. These features are substantial too: an average feature requires almost 900 lines of Chisel code across 14 files, with a maximum of 3000 across 28 files. Interestingly, from a total of 195 iterations, the agent used the \texttt{KEEP} flag 11 times across the translation of 8 features, to override the parent selection mechanism in Sim2RTL.

\begin{table}[t]
\centering
\caption{The 73 translated features by their place in the core. Each feature is placed in the unit that holds most of its RTL change.
\emph{Combinations} are bundles of earlier features of the same part.}
\label{tab:feature-breakdown}
\small
\resizebox{\columnwidth}{!}{
    \begin{tabular}{@{}lr@{\hspace{1.2em}}lr@{\hspace{1.2em}}lr@{}}
    \hline
    \textbf{Frontend} & \textbf{28} & \textbf{Back end} & \textbf{20} & \textbf{Memory subsystem} & \textbf{25} \\
    \thickhline
    Direction prediction & 6  & Rename and recovery      & 6 & LSU and TLB      & 8 \\
    Next-PC and BTB      & 4  & Issue and wake-up        & 3 & L1 data cache    & 7 \\
    Fetch                & 6  & Execution units          & 4 & Miss path and L2 & 7 \\
    I-cache              & 1  & Value pred.\ and reuse   & 3 &                  &   \\
                         &    & Back-end width           & 1 &                  &   \\
    Combinations         & 11 & Combinations             & 3 & Combinations     & 3 \\
    \hline
    \end{tabular}
}
\end{table}

\begin{figure}[t]
    \centering
    \includegraphics[width=\columnwidth]{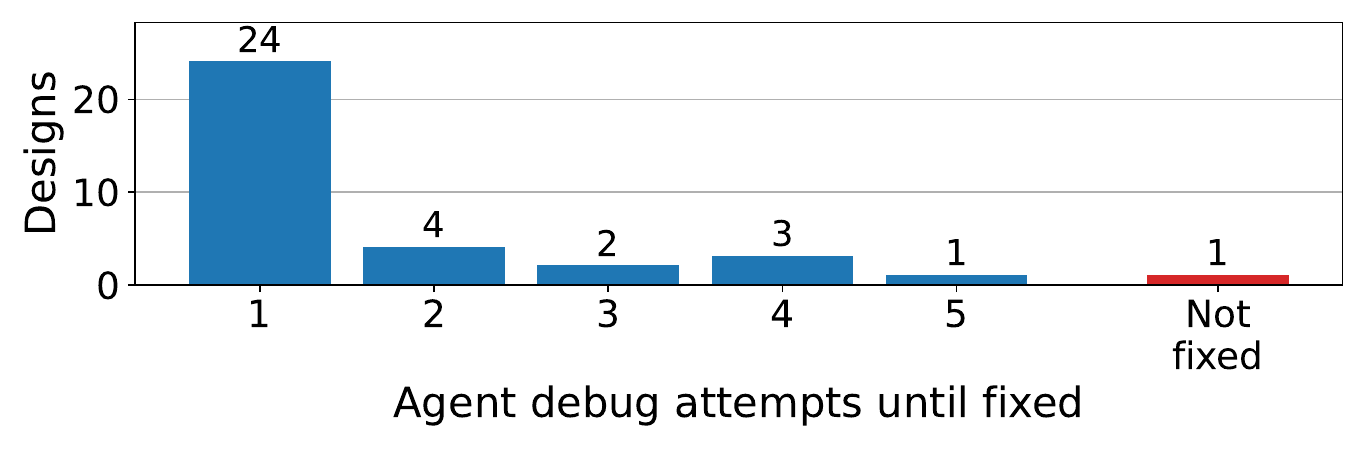}
    \caption{Debug attempts until each failing design was fixed.}
    \label{fig:debug-attempts}
  \end{figure}

Throughout the flow, the lowered RTL runs increasingly complex workloads: from the 87 bare-metal benchmarks of the translation loop to full SPEC17 with reference inputs on FireSim on AWS F2~\cite{F2aws} FPGAs. Bugs were discovered throughout, and all but one discovered bug were fixed by the debugging agent within a budget of 5 iterations. Figure~\ref{fig:debug-attempts} shows a histogram of required debugging iterations on the buggy designs.

Although 35 of the 73 feature implementations initially contained a bug, there are only 26 distinct bugs, 14 of which appear in more than one design, adding up to a total of 72 occurrences. Table~\ref{tab:bug-mechanisms} shows a summary of these bugs. 

Interestingly, two of the bugs were not introduced by the agents at all: they are latent bugs in the off-the-shelf upstream BOOM that the new features exposed. Furthermore, many bugs do not arise from the LLMs repeatedly making the same mistakes when closing gaps in isolation, but rather from mistakes made by agents in prior iterations that are spread through the shared notes.

 \begin{table}[t]
  \centering
  \caption{The 26 distinct bugs by the mechanism they affect. \emph{Occurrences} counts each bug once per feature that contains it.}
  \resizebox{\columnwidth}{!}{
  \small
      \begin{tabular}{l*{2}{>{\raggedleft\arraybackslash}p{1.3cm}}}
      \hline
      Mechanism & Distinct bugs & Occurrences \\
      \thickhline
      Replay of loads that partly overlap a store       & 3  & 33 \\
      L1--L2 coherence handshake (GrantAck, probes)     & 6  & 12 \\
      Early-redirect flush selection                    & 1  & 4  \\
      Branch-killable queue (stock BOOM)                & 2  & 4  \\
      Load wake-up timing                               & 2  & 4  \\
      L1 victim choice and refill bookkeeping           & 3  & 4  \\
      Branch-tag release of checkpointed loads          & 1  & 2  \\
      Dependence-collapse commit hold                   & 1  & 2  \\
      Seven other mechanisms (one bug each)             & 7  & 7  \\
      \thickhline
      Total                                             & 26 & 72 \\
      \hline\\
      \end{tabular}
      }
      \label{tab:bug-mechanisms}
  \end{table}

\subsection{Translation Quality of Results}\label{sec:gem52rtleval:quality}
To close the gap between ideation in these simulators and realistic performance results, we collect the IPC for the full SPEC2017 intspeed suite with reference inputs (SPEC17 ref.), using FireSim, for the first 36 designs that passed the SPEC checkpoints. By doing this we show that the translations are robust enough to successfully execute over 30 trillion instructions of realistic workloads. Throughout the rest of this section, for these 36, we report IPC improvement on SPEC17 ref. from the FPGA simulations. Due to time constraints, we could not run all designs in FireSim, so for the other features we report IPC improvement on the SPEC checkpoints with Verilator~\cite{SnyderVerilator}. For the 36 we ran in FireSim, the performance improvement measured in the SPEC checkpoints correlated well with the improvement measured in SPEC17 ref. 

\begin{figure}[t]
    \centering
    \includegraphics[width=\columnwidth]{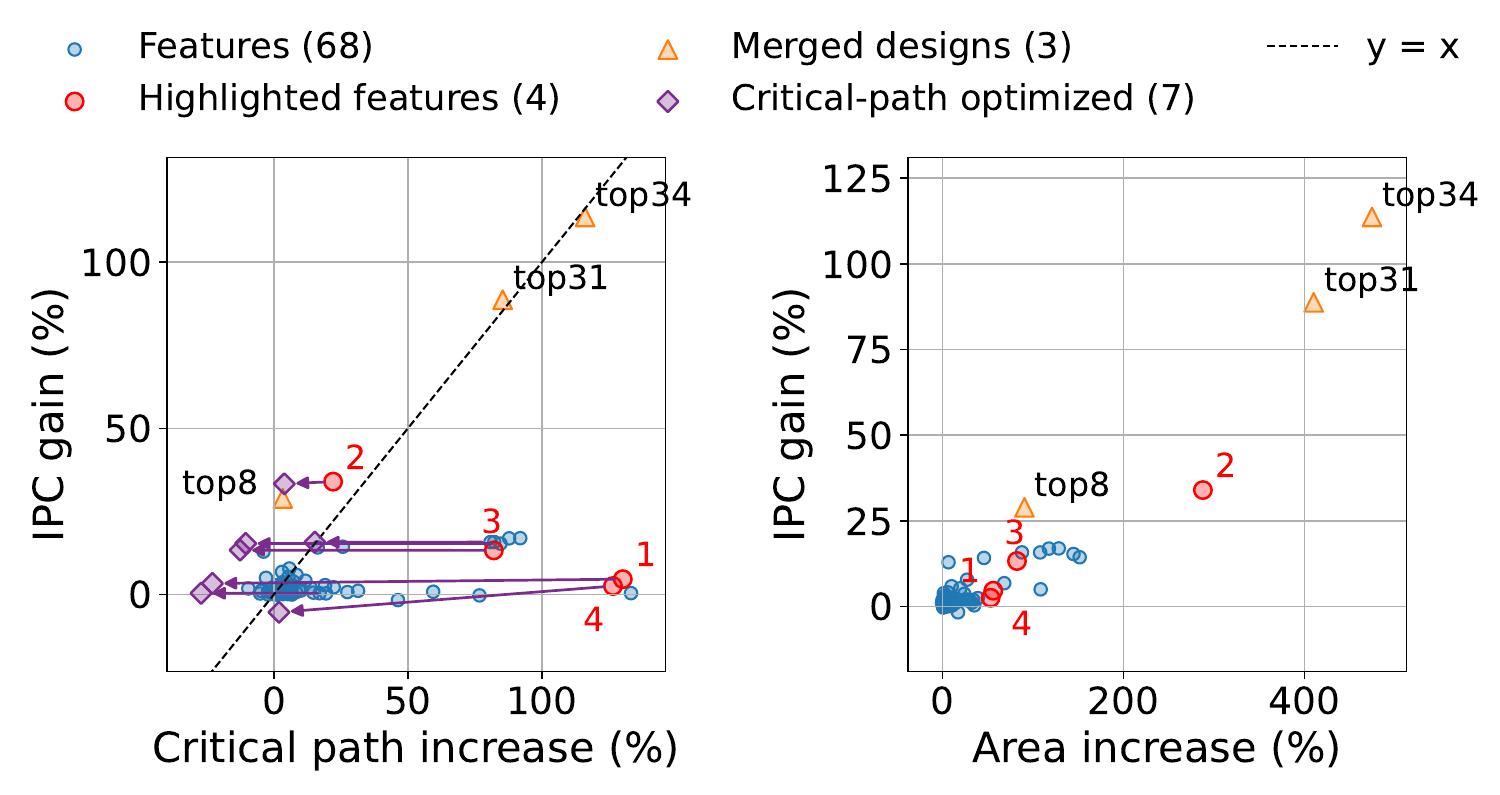}
    \caption{IPC gain against critical path increase (left) and against area increase (right), over the unmodified MegaBOOM, for the 72 features with a synthesis run and the three merged designs. The IPC gain is on SPEC17 ref using FireSim where available (36 features and top 8), and on 11 SPEC17 checkpoints using Verilator otherwise. The left plot leaves out one feature (+0.08\% IPC, +244\% critical path).}

    \label{fig:sky130-scatter}
  \end{figure}

As shown in Figure~\ref{fig:sky130-scatter}, the IPC improvement from the implemented features varies widely from minor reductions in IPC of 1.7\% to improvements of up to 34\%. 

\begin{table}[t]
    \centering
    \caption{Merged designs alignment and SPEC17 speedup over MegaBOOM. Speedups are measured using full SPEC runs on AWS F2 for top 8 and using 143 SimPoint checkpoints for top 31 and top 34.}
    \setlength{\tabcolsep}{4pt}
    \resizebox{0.9\columnwidth}{!}{
        \begin{tabular}{l*{5}{>{\raggedleft\arraybackslash}p{1.3cm}}}
        \hline
         & & \multicolumn{2}{c}{Mismatch factor} & \multicolumn{2}{c}{SPEC17 speedup} \\
        \cmidrule(lr){3-4} \cmidrule(lr){5-6}
        Design & Features & 87 train bmks & 11 CPs & SimPoint 143 CPs & Full SPEC runs \\
        \thickhline
        top8  & 8  & 1.009 & 1.063 & --      & 28.9\% \\
        top31 & 31 & 1.023 & 1.072 & 89.0\%  & --     \\
        top34 & 34 & 1.024 & 1.075 & 113.3\% & --     \\
        \hline\\
        \end{tabular}
        }
        \label{tab:merged-results}
    \end{table}

\begin{figure}[t]
    \centering
    \includegraphics[width=\columnwidth]{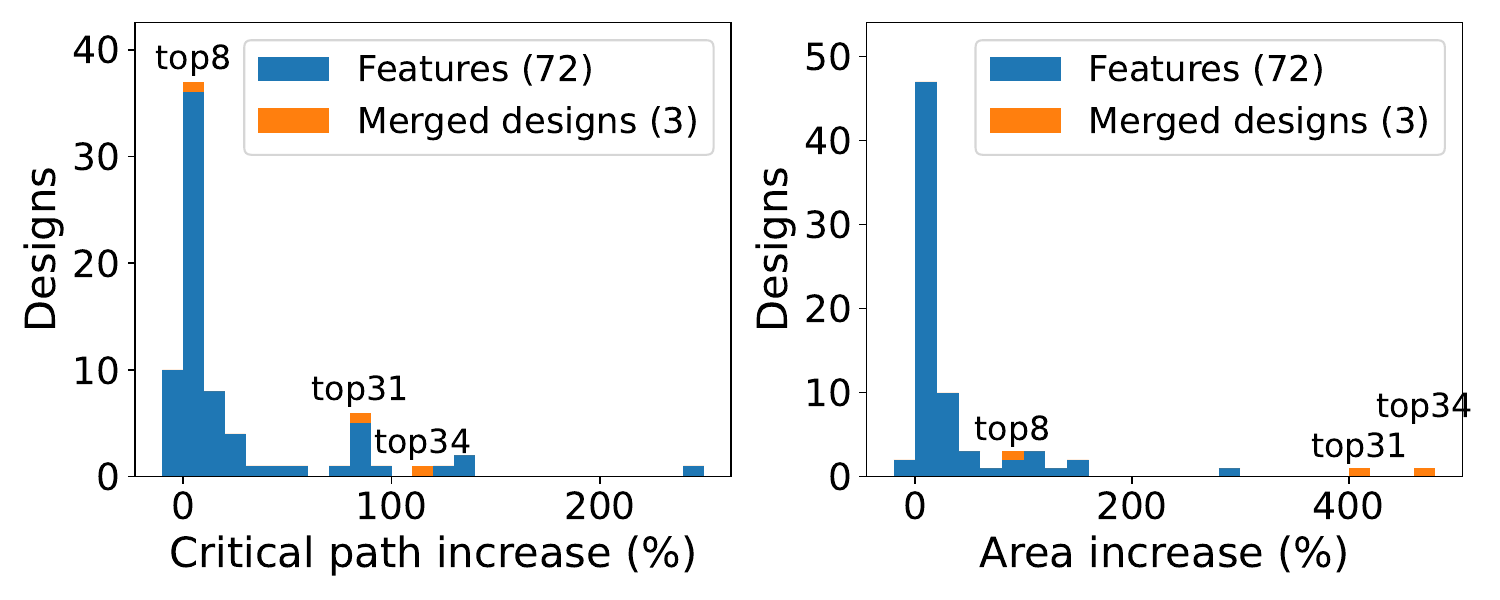}
    \caption{Sky130 critical path (left, 10\% buckets) and area (right, 20\% buckets) increase over the
    unmodified MegaBOOM, for the 72/73 features which are synthesized and the three merged designs. One design timed out after 48 hours in synthesis.}
    \label{fig:sky130-hist}
  \end{figure}

But IPC only tells half of the story. Architects also need area and timing results to fully evaluate their designs. Because every translation is synthesizable RTL, we run gate-level synthesis for all translated features using the SkyWater 130nm PDK~\cite{SKY130} in a commercial synthesis tool (slow corner, SRAMs modeled using CACTI~\cite{muralimanohar2009cacti} scaled to 130nm) and compare them with the unmodified core, which has an area of 31.9~mm$^2$ and critical path of 20.8~ns. 
Figure~\ref{fig:sky130-hist} shows a histogram grouping features into bins based on their corresponding increases in area and timing. Clearly, most designs come with small increases to the area and critical path, but some outliers drastically decrease ASIC quality-of-result (QoR).

Figure~\ref{fig:sky130-scatter} shows two plots: IPC vs. critical path increases on the left and vs. area increases on the right. 
The dashed line on the left plot in Figure~\ref{fig:sky130-scatter} marks where IPC improvements break even with frequency degradation. 
20 out of 72 translated features fall above it, indicating that they were net gains for wall-clock performance.

In contrast with the clear aggregate performance wins, a second bucket of features is realizable but requires a QoR optimization pass to mitigate critical path increases during translation. To demonstrate this, we take 7 optimizations with significant critical path increases, and run the open-source agentic critical path optimization flow packaged with CHIA and described in~\cite{cui26}. The results are shown with arrows in Figure~\ref{fig:sky130-scatter}. Examples of features that fall in this bucket are features 2 and 3. Feature 2 originally included a long path from the L2 cache to the core, so that its responses could wake up dependent instructions in the same cycle. The critical path optimization agent broke this path by adding an extra cycle to L2 responses, which cost only 0.7\% IPC from its original 34\%. Feature 3 adds a statistical corrector on top of TAGE that updates its shared threshold sequentially branch by branch in gem5. In MegaBOOM, this meant 16 serial updates per cycle. The optimization flow modified the RTL to perform the updates in parallel, drastically reducing this path's length with no IPC impact.

A third bucket consists of features which are not physically realizable.  For example, feature 4 decouples the branch predictor from fetch but lets fetch use a prediction in the same cycle it is made; to reduce the critical path, the agent had to remove this same-cycle hand-off, which adds one cycle to every front end restart and costs 7.7\% IPC, three times the feature's 2.5\% gain. Another example is feature 1, which attempted to bring instructions directly from decode to the issue stage in a single cycle, including decoding, load checking, branch tagging, renaming, and dependency collapsing. The optimization flow had to break up this path and the feature lost almost 25\% of its gain (from 4.4\% increase to 3.3\%). 

Note that some features end up with a better critical path than the baseline after running the critical path optimization agent. In these cases the flow also optimized the baseline's critical path.

Area results are similar. Though many features come with small area increases, others drastically increase the size of the core. Many of these blow-ups are the result of large flip-flop arrays, some of which could be changed to SRAMs to reduce area. On the other hand, some implementations could not be made with SRAMs without deviating from the actual implementation reference or sacrificing critical path time and IPC. For instance, many features required multiported 0-cycle reads that must be implemented using flip-flops.

To further test the quality of the final translations we merge the top 8, top 31, and top 34 features together into 3 merged designs, and analyze both their alignment and performance. Table~\ref{tab:merged-results} shows a summary of the results. Compared to an original alignment of 1.048 on the checkpoints, the merged designs maintain a strong alignment, with a maximum mismatch of 1.075 for top34. 
This supports the accuracy of the individual translations given that multiple falsely aligned features could lose alignment when combined. 

Top 8 fits on AWS F2 FPGAs~\cite{F2aws}, so we run it through the full SPEC17 suite with reference inputs (30+ trillion instructions) in FireSim. Top 31 and top 34 do not, so we run the full set of SPEC checkpoints and compute the final IPC using the SimPoint methodology~\cite{Sherwood02} in Verilator. All three yield significant IPC increases. Top 8 shows particularly promising results, with an IPC increase of 28.9\% against a critical path degradation of only 3\%! This result is a strong demonstration of the value of RTL2Sim and Sim2RTL for improving the performance and QoR of real processors.

\subsection{Costs and Scale}
Translating and validating the 73 features required considerable resources, highlighting the scale and complexity of Sim2RTL. This flow has harnessed up to 5056 cores, for a total of 330,301 core-hours. We have also run nearly 10,000 hours of FPGA simulation for validation purposes, running the entire 30 trillion instructions of  SPEC17 ref. on half of the features, and run over 5000 Verilator~\cite{SnyderVerilator} simulations of SPEC checkpoints. 

Our evaluation relied on 624 hours of LLM agent runtime. An average translation cost \$110 including debugging and proposers.

\section{Related Work}\label{sec:relwork}
Misaligned performance models and high-level simulators have long been understood to be a problem in computer architecture~\cite{Skadron03, Gibson00, Nowatzki15Harmful, Carlson14}, and in line with this, building more accurate models and simulators receives significant attention in the community~\cite{Black98, Desikan01, Yi03}. A large number of high-level simulators and performance models have been built, each designed with new techniques to maximize accuracy while aiming for high performance~\cite{Burger97, Vachharajani06, gem5-new, scarab, Bohrer04, Emer02, gober22, Rosenblum97, Pai97, Magnusson02, Martin05, Binkert06, Wenisch06, zsim, Yourst07, Carlson11, Miller10, Rodrigues11, Ahn13, Ryckbosch10}. RTL2Sim is designed to address this problem by aligning these simulators with underlying hardware implementations. Though we focus this work on gem5, RTL2Sim is designed to be retargetable to other simulators, leveraging the flexibility of CHIA~\cite{cui26}.

Prior work has analyzed the accuracy of high-level simulators like gem5~\cite{Butko12, Walker18, Qiu23, Gutierrez14}. For example, Butko et al.~\cite{Butko12} tune a gem5 simulator's parameters to match that of a real dual-core ARM Cortex-A9 SoC (STE Nova A9500), and conclude that the simulator is a decent representation, whereas Gutierrez et al.~\cite{Gutierrez14} modify the gem5 simulator source and parameters to match a Cortex-A15 and conclude that microarchitectural alignment is limited. We find that building a well-aligned simulator is possible, but requires substantial rewriting of the simulator and tuning of configuration parameters to match the underlying design. RTL2Sim addresses the massive engineering effort required to do this.

A lot of work has also gone into methods for tuning and validating parameters of high-level simulators and models~\cite{Adileh19, Huppert21, Ryckbosch10}, including using automated techniques for setting the parameters, such as using Bayesian optimization of gem5's discrete parameters~\cite{Qiu24}, and running gradient descent of a neural network modeling a simulator~\cite{Renda20}. In contrast, our work goes beyond parameter tuning to directly making source code changes to the simulator to introduce missing microarchitectural features and design elements.

Machine learning has also been used to learn performance models for processors~\cite{Li22, Pandey24, Li24, Nasr25}. For example, SimNet~\cite{Li22} trains a neural network to predict the latency of each instruction moving through the core. TAO learns a simulator using functional traces of a program, and uses fine-tuning to support different microarchitectures~\cite{Pandey24}. Unlike RTL2Sim, these works do not produce/learn simulator \textit{code}, but learn a performance model itself. RTL2Sim instead produces an editable simulator, allowing for human architects to leverage the simulator to implement new ideas.

Agentic AI has been used to generate a high-level simulator for a DNN accelerator~\cite{xia26}. Specifically, this work builds a Python simulator of a systolic-array accelerator from natural language descriptions of the needed modules. Our work differs in that the specification for their simulator was natural language, and there is no underlying RTL design to validate the simulator against.

Agentic AI has already been used widely in computer architecture, including for writing RTL~\cite{Deng26, Wu26, Zhang25, zhao24, zhao25, wang25_2, Liu24, Gai25}, designing a complete in-order core~\cite{VerkorTeam26}, and building an LLM inference accelerator~\cite{theverkorteam26_2}. Discovery of novel microarchitectural techniques and structures has also been a popular topic~\cite{Sankaralingam26, Gupta26, Blasberg26}, including new cache replacement policies discovered with evolutionary algorithms~\cite{Gupta26}. This is the first paper to design agentic loops to generate high-level microarchitectural CPU simulators and to translate simulator features to RTL.

\section{Discussion}\label{sec:disc}
One challenge of high-level simulator based computer architecture research is the difficulty in understanding the realizability of an idea in RTL. In particular, it is hard to get a quantitative estimate for how a feature will impact the PPA of an underlying design from just its implementation in a microarchitectural simulator. We saw this clearly in Section~\ref{sec:eval-Gem52RTL}, where the translated features had IPC-accurate RTL implementations, but in multiple cases the area and critical path of the designs increased substantially.

In this vein, our open-source release of the loops developed in this work should provide a path for researchers working in gem5 and other microarchitectural simulators to more easily evaluate the PPA impacts of their new ideas.

Furthermore, because we have automated the translation process, there is an opportunity to improve the performance of open-source hardware designs, by integrating techniques from the literature which have only been implemented in high-level simulators.

\section{Conclusion}\label{sec:conc}
In this paper we described the design of an agentic system for solving two core challenges of computer architecture research: \textbf{(1)} maintaining high-level simulator performance alignment with respect to evolving RTL and silicon implementations of out-of-order superscalar microprocessor designs and \textbf{(2)} automatically translating new ideas implemented in high-level simulators to RTL implementations for ASIC quality-of-result feedback (e.g., area and frequency). We detailed many of the strategies we discovered and lessons we learned while building this system.

To address the first, we designed RTL2Sim, an agentic loop which builds an accurate high-level microarchitectural model for an RTL implementation of a microprocessor, in this case an OoO superscalar design. We used RTL2Sim to build a gem5 simulator with IPC within 3\% of the BOOM core in only 2 days, a huge improvement over the sustained human engineering effort traditionally required to build and maintain such an accurate high-level simulator for a processor.

For the second, we designed Sim2RTL, an agentic loop for translating new microarchitectural features implemented in a well-aligned high-level model into the RTL of a microprocessor design. Sim2RTL successfully translated 72 of 73 attempted features down into the BOOM out-of-order superscalar RISC-V core, at an average of only \$110 of LLM API cost per feature. Each translation matches or improves on the original alignment between gem5 and the RTL on a large suite of 87 benchmarks. Merging 8 features yields a design which improves IPC on the 30 trillion instructions of SPEC17 ref. by 28.9\% for only a 3\% critical path degradation in BOOM.

Furthermore, Sim2RTL demonstrates the importance of considering physical realizability when working in high-level simulators. The features considered fall into three categories: clear wall-clock performance improvements, features which are not realizable, and features which require agentic optimization to turn into improvements. Having Sim2RTL makes it easy and fast to determine which bucket a new idea falls into.

Our agentic loops, generated simulator, and BOOM optimizations will be open-sourced in the near future to enable the broader community to more easily work with high-level microarchitectural simulators and RTL-based implementations together.

\section{Acknowledgements}\label{sec:ack}
The information, data, or work presented herein was funded in part by the Jean and Hing Wong Foundation Faculty Fellowship, Samsung Advanced Institute of Technology (SAIT), and SLICE Lab industrial sponsors and affiliates. Any opinions, findings, conclusions, or recommendations in this paper are solely those of the authors and do not necessarily reflect the position or the policy of the sponsors. This material is also based in part upon work supported by the National Science Foundation Graduate Research Fellowship Program under Grant No. DGE 2146752 and No. DGE 2637800. Any opinions, findings, and conclusions or recommendations expressed in this material are those of the author(s) and do not necessarily reflect the views of the National Science Foundation.

\bibliographystyle{ACM-Reference-Format}
\bibliography{bib}

\end{document}